\documentclass[reprint,aps,prl,twocolumn,notitlepage,nofootinbib,
showpacs,preprintnumbers,superscriptaddress]{revtex4-1}
\usepackage[T1]{fontenc}
\usepackage[utf8]{inputenc}

\usepackage{bm}
\usepackage{amsmath}
\usepackage{amssymb}
\usepackage{esint}
\usepackage{color,xcolor}
\usepackage{graphicx}
\PassOptionsToPackage{normalem}{ulem}
\usepackage{ulem}
\usepackage{times}
\usepackage{hyperref}
\usepackage{amsthm,bm,mathrsfs,stmaryrd,amsmath}
\usepackage{subcaption}
\usepackage{tikz,float}

\newcommand{\be}{\begin{equation}}
\newcommand{\ee}{\end{equation}}
\newcommand{\ba}{\begin{eqnarray}}
\newcommand{\ea}{\end{eqnarray}}

\newcommand{\nn}{\nonumber\\}

\def\pa{\partial}

\def\a{\alpha}

\def\d{\delta}
\def\D{\Delta}
\def\e{\epsilon}

\def\l{\lambda}
\def\L{\Lambda}
\def\m{\mu}

\def\x{\xi}

\def\s{\sigma}

\def\f{\phi}

\def\be{\begin{eqnarray}}
\def\ee{\end{eqnarray}}
\def\D{\Delta}

\def\a{\alpha}

\def\m{\mu}

\def\D{\Delta}
\def\l{\lambda}

\def\d{\delta}
\def\nn{\nonumber\\}
\def\pa{\partial}
\def\s{\sigma}
\def\e{\epsilon}

\newcommand\<\langle
\renewcommand\>\rangle

\begin{document}

\title{Accurate Boundary Bootstrap for the Three-Dimensional O($N$) Normal Universality Class}
\author{Runzhe Hu}
\author{Wenliang Li}
\email{liwliang3@mail.sysu.edu.cn}
\affiliation{School of Physics, Sun Yat-Sen University, Guangzhou 510275, China}

\begin{abstract}
The three-dimensional classical O($N$) model with a boundary has received renewed interest 
due to the discovery of the extraordinary-log boundary universality class 
for $2\leq N< N_c$.  
The critical value $N_c$ and the exponent of the boundary correlation function are related to certain amplitudes in the normal universality class. 
To determine their precise values, 
we revisit the 3d O($N$) boundary conformal field theory for $N=1, 2, 3, 4, 5$. 
After substantially improving the accuracy of the boundary bootstrap, 
our determinations are in excellent agreement with the Monte Carlo results, 
resolving the previous discrepancies due to low truncation orders.  
We also use the recent bulk bootstrap results to deduce highly accurate Ising data.  
Many bulk and boundary predictions are obtained for the first time. 
Our results demonstrate the great potential of  
the $\eta$ minimization method for many unexplored bootstrap problems 
in which positivity constraints are absent. 
\end{abstract}

\maketitle 
\paragraph{Introduction.}As codimension one defects, boundaries are ubiquitous and play an important role in condensed matter physics and high energy physics, 
ranging from edge states of topological materials to D-branes in string theory. 
In this work, we are interested in boundary critical phenomena \cite{Binder:1983,Diehl:1986}. 
The O$(N)$ model with a boundary provides a basic example of boundary criticality,  
which corresponds to the Ising, XY, and Heisenberg universality classes for $N=1,2,3$. 
However, a complete understanding of the three-dimensional  O($N$) boundary phase diagram remains elusive. 

In \cite{Metlitski:2020cqy}, Metlitski pointed out that 
there exists a novel universality class of the extraordinary-log type for $2\leq N<N_c$. 
See \cite{Toldin:2020wbn,Hu:2021xdy,Padayasi:2021sik,Toldin:2021kun,Zhang:2022hpz,Zou:2022mhr, Krishnan:2023cff,Sun:2023vwy,Cuomo:2023qvp,Toldin:2024pqi} for further developments. 
This phase is characterized by {\it logarithmic} decay of the boundary spin correlator 
\be\label{log-correlator}
\<\vec S_{\vec x}\cdot \vec S_{\vec y}\> \sim \frac{1}{(\log|\vec x-\vec y|)^{q}}\,,
\ee 
and its stability hinges on a delicate balance of universal parameters. 
The logarithmic form emerges from a marginally irrelevant coupling of the nonlinear sigma model for the Goldstone modes. 
The exponent $q=(N-1)/(2\pi \a)$ is determined by the universal renormalization group (RG) parameter
\be 
\label{alpha-definition}
\a=\frac1{32\pi} \frac{a_\f^2}{b_{\f t}^2}-\frac{N-2}{2\pi}\,,
\ee
where  $(a_\f, b_{\f t})$ are amplitudes in the normal universality class 
\cite{Binder:1983,Diehl:1986,Bray:1977fvl,Burkhardt:1987,Diehl:1994,Burkhardt:1994,Diehl:1994-2}.  
The extraordinary-log phase is stable for $\a>0$. 
The critical value $N_c$ is defined by $\a|_{N=N_c}=0$. 
Above $N_c$, the extraordinary fixed point may annihilate with the special fixed point or evolve into a power-law type \cite{Metlitski:2020cqy}.  
For $N=2,3$, these amplitudes were extracted from Monte Carlo simulations \cite{Toldin:2021kun}. 
Their precise values at larger $N$ are crucial for the determination of $N_c$, 
but Monte Carlo results are not yet available. 
The previous conformal bootstrap study \cite{Padayasi:2021sik} provided important estimates for a larger range of $N$. 
In conformal field theory (CFT), 
the universal amplitudes $(a_\f, b_{\f t})$ are known as boundary operator expansion (BOE) coefficients. 
Due to the presence of an explicit symmetry breaking field, 
a normal boundary partly breaks not only the external conformal symmetry, 
but also the internal O($N$) symmetry.   
Thus, the normal boundary CFT (BCFT) is of interest on its own,    
and we also study the Ising case of $N=1$.  

\begin{figure}
\centering
\begin{tikzpicture}[scale=0.7]

\draw [thick] (-4.5,-1) -- (-1.5,-1);
\node at (-5.5,0) {\scalebox{1.2}{$\displaystyle\sum_k$}};
\draw [thick] (-3,0) -- node [anchor=east] {$\mathcal{O}_k$} (-3,-1) ;
\draw [thick] (-3,0) -- (-4,1) node [anchor=east] {$\mathcal{O}_1$};
\draw [thick] (-3,0) -- (-2,1) node [anchor=west] {$\mathcal{O}_2$};
\node at (1.5,0) {\scalebox{1.2}{$\displaystyle\sum_n$}};
\draw [thick] (2.5,-1) --  (5.5,-1);
\draw [thick] (3,-1) -- (3,1) node [anchor=east] {$\mathcal{O}_1$};
\draw [thick] (5,-1) -- (5,1) node [anchor=west] {$\mathcal{O}_2$};
\node at (4,-0.6) {$\hat{\mathcal{O}}_n$};
\node at (0,0) {$\scalebox{1.5}{=}$};
\end{tikzpicture}
\caption{Crossing symmetry of $\langle \mathcal{O}_1\,\mathcal{O}_2\rangle$ in boundary CFT.}
\label{crossing symmetry}
\end{figure}
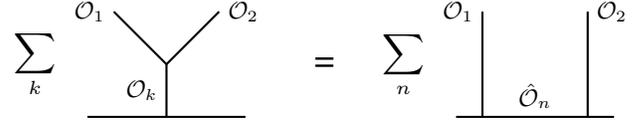

Since the seminal work \cite{Rattazzi:2008pe}, 
the conformal bootstrap program for  $d>2$ CFT has been revived by incorporating positivity constraints 
and efficient algorithms \cite{Poland:2018epd,Rychkov:2023wsd}. 
See \cite{El-Showk:2012cjh,Kos:2014bka,El-Showk:2014dwa,Simmons-Duffin:2015qma, Kos:2013tga,Kos:2015mba,Kos:2016ysd,Simmons-Duffin:2016wlq,Chester:2019ifh, Chester:2020iyt, Reehorst:2021hmp, Chang:2024whx} 
for the impressive progress on the 3d O($N$) bulk data  based on the positive bootstrap. 
In \cite{Liendo:2012hy}, 
Liendo, Rastelli, and
van Rees extended the bootstrap program 
to BCFT. 
However, it is not clear if the powerful positive bootstrap is applicable,
as the bulk-channel expansion of a two-point correlator is not quadratic. 
(See the left-hand side of Fig.\ref{crossing symmetry}.) 
This necessitates alternative, non-positivity-based methods. 

In \cite{Gliozzi:2013ysa}, Gliozzi proposed to solve the bootstrap equation by truncating to 
a finite number of operators, which does not rely on positivity. 
In \cite{Gliozzi:2013ysa, Gliozzi:2014jsa, Esterlis:2016psv}, 
the truncated bootstrap constraints on operator dimensions are encoded in 
determinants or singular values. 
While early applications to the O($N$) BCFT in  \cite{Gliozzi:2015qsa,Gliozzi:2016cmg, Padayasi:2021sik} were promising,   
these studies were limited to low truncation orders.  
For the normal universality class, 
the previous bootstrap studies \cite{Gliozzi:2015qsa,Padayasi:2021sik} yielded results 
that were in noticeable tension with 
Monte Carlo estimates \cite{Toldin:2021kun,Przetakiewicz:2025gzi}, 
casting doubt on the reliability of the truncated bootstrap approach. 
As the truncated bootstrap is not a rigorous method, 
the tension with \cite{Padayasi:2021sik} is not a mathematical contradiction. 
It was not clear if these discrepancies arose from uncontrolled systematic errors 
or low truncation orders. 
To address this question, we use the $\eta$ minimization formulation of the truncated bootstrap \cite{Li:2017ukc},  
which also has a number of variants due to its flexibility, 
incorporating artificial intelligence \cite{Kantor:2021kbx,Kantor:2021jpz,Kantor:2022epi,Niarchos:2023lot,Niarchos:2025cdg}, 
analytic input \cite{Li:2023tic,Poland:2023bny,Barrat:2025wbi,Poland:2025ide}, 
and random weights \cite{Poland:2023vpn, Poland:2023bny,Barrat:2025wbi,Poland:2025ide}.

By reformulating the truncated boundary bootstrap as a search for the zeros of a cost function, 
we reach significantly higher truncation orders $\Lambda$  
(see  Table \ref{MaxLambda}) 
and substantially improve the accuracy of the bootstrap results.  
We introduce a new procedure to construct starting points for local minimizations 
and substantially enhance the efficiency of extracting sparse solutions.  
(Sparseness in the truncated bootstrap is the counterpart of positivity in the conventional bootstrap.)
Then, the bootstrap estimates are in excellent agreement with the Monte Carlo results, 
resolving the previous discrepancies. 
Many results are more accurate or completely new. 
Furthermore, we assign reliable errors to the estimates associated with low-lying operators, 
which are mainly from uncertainties in bulk input parameters.   
To obtain highly accurate Ising results, 
all we need are precise bulk input and high enough truncation orders.

Since the truncated bootstrap was originally devised for nonunitary CFTs in \cite{Gliozzi:2013ysa}, 
we anticipate that our substantially improved version will overcome long-standing technical bottlenecks  
and find wide application in statistical mechanics, from random geometry to disordered systems. 
These systems involve more intricate theoretical descriptions, such as logarithmic CFTs and supersymmetric nonlinear sigma models.

\begin{table}[t]\centering
\begin{tabular}{cccccc}
\hline
\hline
& $N=1$ & $N=2$  & $N=3$& $N=4$ & $N=5$\\
\hline
This work~  &88, 68& 76   &68 & 60 & 88 \\
\hline
\cite{Padayasi:2021sik,Gliozzi:2015qsa} &9, 8& 10  &10 & 9 & 9 \\
\hline
\end{tabular}
\caption{The maximum truncation order $\L_{\text{max}}$ in this work and the previous truncated bootstrap studies \cite{Padayasi:2021sik,Gliozzi:2015qsa} . 
The two numbers for $N=1$ are associated with $\langle \s\,\e\rangle$ and $\langle \s\,\s\rangle$.  
}
\label{MaxLambda}
\end{table}

\paragraph{Boundary bootstrap with the $\eta$ minimization.}According to boundary conformal symmetry, the correlation function of two bulk scalar primaries reads
\be\label{2-pt}
\<\mathcal O_1(x)\,\mathcal O_2(y)\>=\frac{G(\x)}{(2x_\perp)^{\D_1}(2y_\perp)^{\D_2}}\,\x^{-\frac{\D_1+\D_2}{2}}\,,
\ee
where $G(\x)$ is an unknown function of the conformally invariant cross ratio $\x=(x-y)^2/(4x_\perp y_\perp)$. 
We can decompose the two-point function, Eq. \eqref{2-pt}, into conformal blocks. 
In the bulk channel, we use the bulk operator product expansion (OPE)
\be
\mathcal O_1(x)\, \mathcal O_2(y)
=\sum_k\l_{12k}\,C_{12k}(x-y,\pa_y)\,\mathcal O_k(y)\,,
\ee
so Eq. \eqref{2-pt} is given by a summation of bulk one-point functions,
\be
\<\mathcal O_k(x)\>=\frac{a_k}{(2x_\perp)^{\D_k}}
\ee 
and their derivatives, 
where $a_k$ vanishes for spinning primaries. 
In the boundary channel, we consider the boundary operator expansion (BOE) or the bulk-to-boundary OPE
\be
\mathcal O_k(x)=\sum_{n}b_{kn}\, \hat C_{kn}(x_\perp,\pa_{x_\parallel})\,\hat{\mathcal O}_n(x_\parallel)\,;
\ee
then, Eq. \eqref{2-pt} becomes a summation of boundary two-point functions. 
The above one-point coefficients are the BOE coefficients associated with the boundary identity, i.e., $a_{k}=b_{k I}$. 
The agreement between the two decompositions implies the bootstrap or crossing equation
(see also Fig. \ref{crossing symmetry})
\be\label{bdy-crossing}
\sum_k \l_{12k}a_k\, f_{\D_k}^{\D_{12}}(\x)
-\x^{\frac{\D_1+\D_2}{2}}\sum_n \m_{12n}\,\hat f_{\hat\D_n}(\x)=0\,,
\ee
where $\D_{12}=\D_1-\D_2$ and $\m_{12n}=b_{1n}b_{2n}$. 
The descendant contributions in $C(x-y,\pa_y)$ and $\hat C(x_\perp,\pa_{x_\parallel})$ are repackaged as the bulk and boundary channel conformal blocks \cite{McAvity:1995zd}:
\be
f_{\D}^{\D_{12}}(\x)
&=&\x^{{\D}/{2}}{}_2F_1
\Big[ \frac {\D+\D_{12}}{2},\frac{\D-\D_{12}}{2};\D-\frac{d-2}{2}; -\x
\Big]\,,
\nn
\hat f_{\hat\D}(\x)&=&
\x^{-\hat\D}\,
{}_2F_1
\Big[
\hat\D,\hat\D-\frac {d}{2}+1;2\hat\D-d+2; -{1}/{\x}
\Big]\,.\quad
\ee
As there is only one cross ratio, 
it is simpler to study Eq. \eqref{bdy-crossing} than 
the bulk four-point bootstrap equation without a boundary. 
In the bulk channel, we make use of the previous accurate determinations of the bulk operator dimensions. 
In the boundary channel, 
the leading operators in the normal universality class have protected dimensions.  
Together with sparseness of the low-lying spectra, the normal boundary universality class is a natural target for the conformal bootstrap \cite{Padayasi:2021sik}.

\begin{figure}[t]
\centering
\includegraphics[width = 1\linewidth]{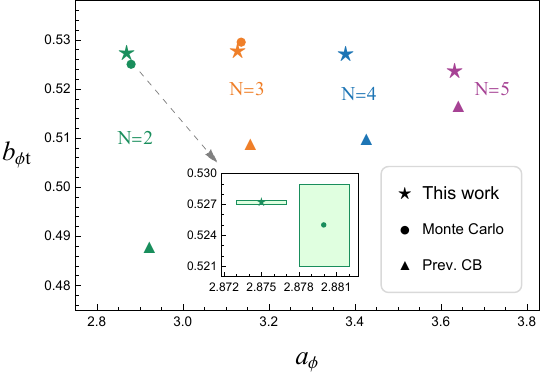}
\caption{The $N=2,3,4,5$ estimates for the universal amplitudes $(a_{\f},b_{\f t})$. 
Our results (stars) show excellent agreement with the Monte Carlo estimates (dots) for $N=2,3$ \cite{Toldin:2021kun}  and resolve the discrepancies seen in the previous conformal bootstrap (CB) study \cite{Padayasi:2021sik} (triangles). The inset shows the $N=2$ error bars from this work and \cite{Toldin:2021kun}.   
}
\label{bphit-aphi-N2345}
\end{figure}

To discretize the bootstrap equation, Eq. \eqref{bdy-crossing}, 
we take the $m$th derivative with respect to $\x$ and then set $\x=1$. 
We restrict the order of $\x$ derivatives to $M$, so we have a finite system. 
We truncate the bulk OPE and BOE in Eq. \eqref{bdy-crossing}, i.e., $k=1,2,\dots,n_\text{bulk}$ and $n=1,2,\dots, n_\text{bdy}$. (See Supplementary Material \cite{SM} for more details.) 
We use the $\eta$ function to encode these truncated bootstrap constraints
\be\label{eta-fn}
\eta=\sum_{j}\sum_{m=0}^{M_j}\left|\pa_\x^m\left(\text{bootstrap equation}_j\right)\right|^2_{\x=1}\,,
\ee
where $j$ labels the bootstrap equations under consideration. 
We impose that the number of bootstrap constraints is the same as that of free parameters, 
which is also referred to as the truncation order $\L$. 
By construction, the $\eta$ function can vanish only when all the truncated bootstrap equations are satisfied. 
Below, we systematically solve the truncated bootstrap equations 
by searching for the zeros of the $\eta$ function, Eq. \eqref{eta-fn},
\be
\eta\big(\{\D_i, \l_{ijk} a_k\}, \{\hat \D_n, b_{in}\}\big)=0\,,
\ee
which are equivalent to the intersection points of certain vanishing loci of minors in Gliozzi's determinant formulation.

\paragraph{The O($N$) BCFT for $N=2,3,4,5$.}In the normal transition, 
the O($N$) symmetry  is explicitly broken to O($N-1$). 
While the bulk operators are classified by O($N$) irreducible representations, 
the boundary operators are associated with O($N-1$). 
The two-point function of the lightest bulk O($N$) vector $\phi_a$ involves two O($N-1$) singlets, 
so we have two crossing equations. 
They correspond to 
\be
\<\phi_N(x)\,\phi_N(y)\>\,,\quad
\sum_{i}\<\phi_{i}(x)\,\phi_{i}(y)\>\,,
\ee
where $a=(i,N)$ and $i=1,2,\dots, N-1$. 
The bulk fusion rule for the product of two O($N$) vectors reads
\be
\phi_a\times \phi_b\sim 
\sum_S\d_{ab}\,\mathcal O
+\sum_T\mathcal O_{(ab)}
+\sum_A\mathcal O_{[ab]}\,,
\ee
which involves the O($N$) singlets $S$, traceless symmetric tensors $T$, and antisymmetric tensors $A$. Only the first two types of representations can be scalar primaries and contribute to the boundary bootstrap equations. 
The boundary fusion rules associated with the O($N-1$) singlets $\hat S$ and vectors $\hat V$ are
\be\label{boundary-fusion}
\phi_N\sim1+  D+\sum_{\hat \Delta>3}\hat{\mathcal O}^{(\hat S)}\,,\quad
\phi_i\sim t_i+\sum_{\hat \Delta>2}\hat{\mathcal O}_i^{(\hat V)}\,,\quad
\ee 
where $D$ is the displacement operator with $\hat\D_{ D}=3$, and $ t_i$ is the tilt operator with $\hat\D_{ t}=2$. 
See Eqs. (2.20) and (2.21) in \cite{Padayasi:2021sik} for the explicit crossing equations. 
The input parameters are the bulk dimensions $\{\D_\phi,\D_{S},\D_{S^{\prime}},\D_{T}\}$ from 
Monte Carlo simulations \cite{Hasenbusch:2019jkj, Hasenbusch:2020pwj,Hasenbusch:2021rse,Hasenbusch:2025yrl} 
and the bulk bootstrap \cite{Kos:2013tga, Chester:2019ifh,Chester:2020iyt}. 
The operators in the fusion rules are ordered by scaling dimensions.  
The subleading ones are indicated by primes. 
For instance, the subleading O($N$) singlet is denoted by $S'$.

Using the $\eta$ minimization method, we systematically increase the truncation order $\L$. 
Surprisingly, the bootstrap results exhibit nice convergence patterns,  
so we further make some {power-law} fits in $1/\L$ and  extract the $\L\rightarrow\infty$ extrapolations. 
We assume that the truncation errors vanish in the infinite $\L$ limit,  
so the errors are from the uncertainties in the bulk input and the $\L\rightarrow\infty$ extrapolations.  
In Fig. \ref{bphit-aphi-N2345}, we compare our estimates for $(a_{\f}, b_{\phi t})$ 
with the literature results. 
As substantial improvements of the previous results in \cite{Padayasi:2021sik}, 
we significantly increase the bootstrap accuracy, 
and our results are in excellent agreement with the Monte Carlo estimates \cite{Toldin:2021kun}. 

\begin{figure}[t]
\centering
\includegraphics[width = 1\linewidth]{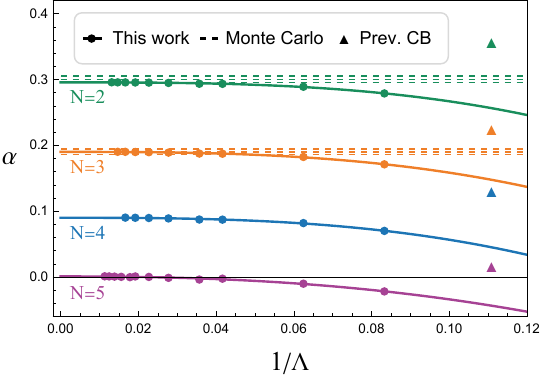}
\caption{The universal RG parameter $\a$ in Eq. \eqref{alpha-definition} at various truncation orders $\L$ (dots). 
For comparison, we also plot the Monte Carlo results with errors \cite{Toldin:2021kun} (dashed lines) 
and the previous conformal bootstrap results \cite{Padayasi:2021sik}  (triangles, $\L=9$). 
}
\label{alpha-N2345}
\end{figure}

In \cite{Padayasi:2021sik}, the $T$ contributions in the bulk channel are projected out 
by considering a linear combination of the two crossing equations. 
Here, we take into account the $T$ contributions and 
solve two crossing equations simultaneously. 
This difference is essential to the convergence of the bootstrap results, 
and explains why the triangle points in Fig. \ref{alpha-N2345} 
lie far from the extrapolation curves. 
In Fig. \ref{alpha-N2345}, we present our results for $\a$ at various $\L$, 
which allow for simple power-law fits. 
The infinite $\L$ extrapolations are well consistent with the Monte Carlo results \cite{Toldin:2021kun}. 
On the other hand, the singlet projection leads to unstable, nonconvergent results. 
(See Fig. 1 of \cite{SM}.) 
They do not follow the trend of the Fig. \ref{alpha-N2345} curves 
because the presence of two boundary sectors 
leads to a less sparse spectrum, which is less suitable for the truncated bootstrap.

For $N=5$, the sign of $\a$ is important for determining the critical value $N_c$. 
Previously, the boundary bootstrap study \cite{Padayasi:2021sik} obtained a result for $\a|_{N=5}$ around $0.017$. 
Without assigning error, it is not clear if $\a|_{N=5}$ is really positive, i.e., if $N_c$ is above $5$. 
Our estimate for $\a|_{N=5}$ is 1 order of magnitude smaller, but still marginally positive.
Based on the estimates $\a|_{N=2,3,4,5}$, 
we use {power-law} fits to extract the critical value
\be
N_c^\text{this work}=5.016(28)\,,
\ee 
which is surprisingly close to the integer value $5$. 
For comparison, a fit of the previous results in \cite{Padayasi:2021sik} gives $N_c\approx 5.2$.

The use of two crossing equations also allows us to determine  
the bulk operator dimensions of the subleading O($N$) traceless-symmetric tensors. 
Using linear fits, we find
\be
\D_{T'}={3.6484(24)}\,,\, 3.559(4)\,,\, 3.49(3)\,,\, 3.354(17)\,,\quad
\ee
for $N=2,3,4,5$.  
Our $N=2$ estimate is in nice agreement with the bulk bootstrap result $3.650(2)$ in  
\cite{Chester:2019ifh}. 
The results for $N=3,4,5$ are new.  
We also obtain rough estimates for the subleading boundary dimensions, $(\hat\D_{\hat S}, \hat\D_{\hat V})\sim 5$. 
Using the bulk OPE coefficients $(\l_{\f\f S}, \l_{\f\f T})$ in \cite{Chester:2019ifh,Chester:2020iyt}, 
we further deduce the following one-point coefficients for the first time:
\be 
N=2:&\quad a_{S}=5.571(12)\,,&\quad \,a_{T}=3.897(6)\,,\\
N=3:&\quad a_{S}=5.369(12)\,,&\quad \,a_{T}=8.406(12)\,. 
\ee

\paragraph{The Ising BCFT.}Using the $\eta$ minimization, we can systematically increase the truncation orders,  
so the accuracy is mainly limited by the bulk input. 
To demonstrate this more clearly, 
we leverage the unprecedentedly precise determinations of the bulk Ising data in \cite{Chang:2024whx}:
\be\label{Ising-input}
\D_\s^{\text{input}}&=&0.518148806(\textbf{24})\,,\,
\D_\e^{\text{input}}=1.41262528(\textbf{29})\,, \qquad\\
\l_{\s\s\e}^{\text{input}}&=&1.05185373(11)\,.  \label{lambda-sse}
\ee

In the normal transition, the $\mathbb{Z}_2$ symmetry of the Ising universality class is broken. 
The boundary fusion rules read
\be
\s,\e\sim 1+ D+ \hat{N}+\hat{N}'+\dots\,.
\ee
We consider two crossing equations in the Ising BCFT.  
The first one concerns the mixed spin-energy correlator $\<\s(x)\,\e(y)\>$,
which is nonvanishing due to symmetry breaking.   
The bulk fusion rule is
\be
\s\times \e\sim \s+\s'+\s''+\dots\,,
\ee
so there are only $\mathbb{Z}_2$-odd operators. 
The dimension of the leading irrelevant operator is about $\D_{\s'}\approx 5.3$. 
As the bulk spectrum has a large gap \cite{Recombination}, 
we expect to obtain highly accurate results. 
The second crossing equation is about the spin-spin correlator $\<\s(x)\,\s(y)\>$,  
which is related to the bulk fusion rule
\be
\s\times \s\sim 1+\e+\e'+\e''+\dots\,.
\ee
The scaling dimension of ${\e'}$ is also an input parameter. 
We mainly use the rigorous result $\D^\text{input}_{\e'}=3.82951(\textbf{61})$ in  \cite{Reehorst:2021hmp}. 
The uncertainty in $\D_{\e'}$ is the main source of error in the Ising boundary bootstrap. 
For this reason, we solve the two crossing equations separately 
and choose a larger maximum truncation order $\L_\text{max}$ for the $\<\s\,\e\>$ crossing equation. (See Table \ref{MaxLambda}.)

\begin{figure}[t]
\centering
\includegraphics[width = 0.95\linewidth]{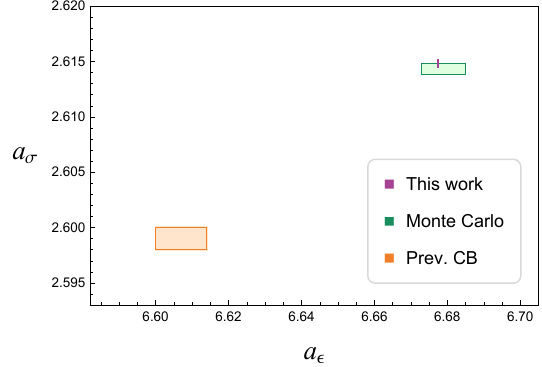}
\caption{The 3d Ising one-point coefficients $(a_{\s},a_{\e})$ from 
this work (purple), Monte Carlo simulations  \cite{Przetakiewicz:2025gzi} (green), and the previous conformal bootstrap \cite{Gliozzi:2015qsa} (orange). 
The uncertainty in our estimate for $a_\e$ is smaller than the symbol size. 
}
\label{asigma-aepsilon-N1}
\end{figure}

Again, we observe nice convergence patterns as $\L$ grows. 
We also use the power-law fits to deduce the $\L\rightarrow \infty$ extrapolations. 
In Fig. \ref{asigma-aepsilon-N1}, we compare our estimates for 
the one-point coefficients of the bulk relevant operators 
with the literature results. 
Our accurate results,
\be
a_\s=2.6148(3)\,,\quad
a_\e=6.677424(16)\,,
\ee 
are in excellent agreement with the Monte Carlo results \cite{Przetakiewicz:2025gzi} 
and resolve the previous discrepancies due to low truncation orders in  \cite{Gliozzi:2015qsa}. 
Using the bulk OPE coefficients from \cite{Reehorst:2021hmp,Simmons-Duffin:2016wlq}, 
we obtain the new one-point coefficients
\be
a_{\e^{\prime}}=42.46(14)\,,\; a_{\s'}={110(4)}\,,\; a_{\e''}=267.8(15)\,.\quad
\ee
We estimate some boundary dimensions and BOE coefficients:
\be
\hat{\D}_{\hat{N}}=5.8792(12)\,,\quad \hat{\D}_{\hat{N}^{\prime}}&=&8.086(24)\,,\\
b_{\e \hat N}=0.2147(23)\,,\quad b_{\e \hat{N}^{\prime}}&=&0.0464(42)\,,\\
b_{\s \hat{N}}=0.00946(11)\,,\quad b_{\s \hat{N}^{\prime}}&=&0.00130(12)\,. 
\ee
The operator dimensions are derived from $\langle \s\,\e\rangle$ due to smaller input uncertainties. 
Our estimate for $\hat{\D}_{\hat{N}}$ is consistent with 
the first two digits of the fuzzy sphere result, $5.858$, in \cite{Dedushenko:2024nwi}. 

As a test of our error analysis, we extract the bulk OPE coefficient $\l_{\s\s\e}$ from our boundary bootstrap results:
\be
\l_{\s\s\e}^{\text{this work}}=1.05184(13)\,.
\ee
The impressive agreement with the bulk bootstrap result, Eq. \eqref{lambda-sse}, 
suggests that our errors are reliable for low-lying operators. 
Another nontrivial check comes from the Ward identity associated with the displacement operator $D$. 
In a given boundary universality class, 
the Zamolodchikov norm of $D$,  
\be
C_ D=\left(\frac{\D_{\mathcal O}\,{a_{\mathcal O}}{}}{4\pi\,b_{\mathcal O D}}\right)^2
=
\begin{cases}
0.18966(10)\quad \mathcal O=\s
\\
 0.18970(10)\quad \mathcal O=\e\,,
\end{cases}
\ee
should not depend on the choice of the bulk operator $\mathcal O$ \cite{Toldin:2021kun,Cardy:1990xm}. 
The difference is consistent with error estimates and much less than that in \cite{Gliozzi:2015qsa}. 
Our results are roughly compatible with the Monte Carlo results, $0.193(5)$ in \cite{Toldin:2021kun}
and $0.198(3)$ in \cite{Przetakiewicz:2025gzi}.

\paragraph{Discussion.}In summary, we resolved the previous discrepancies 
between the conformal bootstrap and Monte Carlo studies of 
the 3d O($N$) normal universality class. 
By implementing the $\eta$ minimization that allows for a systematic increase in the truncation order $\L$, 
we demonstrated that the earlier disagreements were artifacts of low truncation orders, not fundamental limitations of the truncated bootstrap approach. 
The truncated bootstrap results achieved excellent agreement with Monte Carlo benchmarks
and are superior in several aspects. 
Some estimates are two orders of magnitude more accurate. 
Our bulk results also agree well with those from the conventional bootstrap method.
Many bulk and boundary data are completely new.  
We obtain an accurate estimate for  
the critical value $N_c$, which is surprisingly close to $5$.

It would be interesting to consider larger bootstrap systems, 
i.e., correlators of higher points, boundary operators, and other bulk operators. 
A promising future direction is to apply the $\eta$ minimization method to other nonperturbative defect bootstrap problems \cite{Liendo:2012hy,Billo:2016cpy,Lauria:2017wav}. 
It is also important to bootstrap nonunitary CFTs, 
which describe the critical behavior of 
the O($N$) loop model with noninteger or nonpositive $N$,   
the Yang-Lee edge singularity, percolation, and disordered systems 
\cite{Gliozzi:2013ysa,Gliozzi:2014jsa,Shimada:2015gda, Hogervorst:2016itc,Hikami:2017hwv,Hikami:2017sbg,Leclair:2018trn,Hikami:2018qpz,Padayasi:2023hpd}. 
When positivity violations are significant, 
the positive bootstrap methods may not be applicable 
even for the bulk crossing equation.  
But we can still use the truncation methods, such as the $\eta$ minimization. 
It should also be helpful to revisit the bulk bootstrap of the standard Ising model, so our improved Gliozzi method can be tested against a known case.
The convergent patterns were observed in other truncated bootstrap studies \cite{Li:2023tic, Li:2024ggr}, 
which seem to be a general feature and  
should be useful to many unexplored nonpositive bootstrap targets. 

\bigskip
\paragraph{Acknowledgments.}We would like to thank Yongwei Guo, Ning Su, and Shuai Yin for helpful discussions. 
We also thank the anonymous referees for their constructive comments.
This work was supported by the Natural Science Foundation of China (Grants No. 12522504 and No. 12205386).

\paragraph{Data availability.}The data that support the findings of this article are openly available \cite{Hu:2026Zenodo}.

 \bigskip
  \bigskip
\onecolumngrid
 \parbox[c][3em][t]{\textwidth}{\centering \large\bf End Matter}
 
\twocolumngrid
\paragraph{More details about the $N=2,3,4,5$ results.}In Table \ref{Res-2345}, we list some of our boundary bootstrap results for $N=2,3,4,5$  
and some literature results for comparison.
While the accuracy of $a_{\f}$  is comparable to that of the Monte Carlo results for $N=2,3$, 
our bootstrap estimates for $b_{\f t}$ and $\a$ appear to be more accurate. 
Our new bootstrap result for $N=4$ is also more compatible with the unpublished Monte Carlo result \cite{alpha-N=4}, $\a|_{N=4}=0.097(3)$, mentioned in \cite{Toldin:2024pqi}. 

\begin{table}[t]\centering
$N=2$\\
\begin{tabular}{clllll}
\toprule
\multicolumn{1}{c}{Method} & \multicolumn{1}{c}{$a_{\f}$} & \multicolumn{1}{c}{$b_{\f t}$} & \multicolumn{1}{c}{ $b_{\f D}$}& \multicolumn{1}{c}{$\a$}  \\
\hline
This work & 2.875(2)  & 0.5272(2) &0.2440(4)& 0.2957(6)    \\
MC \cite{Toldin:2021kun} & 2.880(2)    &  0.525(4) & & 0.300(5)    \\
CB \cite{Padayasi:2021sik}  & 2.923   &   0.4882   & 0.2701& 0.3567     \\
\hline
\end{tabular}\bigskip

$N=3$ \\
\begin{tabular}{clllll}
\hline\hline
\multicolumn{1}{c}{Method} & \multicolumn{1}{c}{$a_{\f}$} & \multicolumn{1}{c}{$b_{\f t}$} & \multicolumn{1}{c}{ $b_{\f D}$}& \multicolumn{1}{c}{$\a$} \\
\hline
This work & 3.129(3)   & 0.5278(2) & 0.2406(6)&0.1903(7) \\
MC \cite{Toldin:2021kun} & 3.136(2)    &  0.529(3)  & &  0.190(4)      \\
CB \cite{Padayasi:2021sik} &    3.159 &   0.5092     &0.2690 & 0.2236   \\
\hline
\end{tabular}\bigskip

$N=4$ \\
\begin{tabular}{clllll}
\hline\hline
\multicolumn{1}{c}{Method} & \multicolumn{1}{c}{$a_{\f}$} & \multicolumn{1}{c}{$b_{\f t}$} & \multicolumn{1}{c}{ $b_{\f D}$}& \multicolumn{1}{c}{$\a$}  \\
\hline
This work & {3.380(7)}   & 0.5272(13) &0.2369(24)&0.0906(35)    \\
CB \cite{Padayasi:2021sik} &    3.429 &   0.5105   &  0.2758 & 0.1304 \\
\hline
\end{tabular}\bigskip

$N=5$ \\
\begin{tabular}{clllll}
\hline\hline
\multicolumn{1}{c}{Method} & \multicolumn{1}{c}{$a_{\f}$} & \multicolumn{1}{c}{$b_{\f t}$} & \multicolumn{1}{c}{ $b_{\f D}$} & \multicolumn{1}{c}{$\a$}  \\
\hline
This work & 3.634(5)   & 0.5235(5) &0.2390(9) & 0.002(2)  \\
CB \cite{Padayasi:2021sik} &    3.641 &    0.5166    & 0.2653 & 0.0166 \\
\hline
\end{tabular}
\caption{Some $N>1$ BOE coefficients and the universal RG parameter $\a$  
from this work, Monte Carlo (MC) simulations \cite{Toldin:2021kun},  
and the previous conformal bootstrap study \cite{Padayasi:2021sik}. 
}
\label{Res-2345}
\end{table}

\begin{table}[t]\centering
\begin{tabular}{clllll}
\hline\hline
\multicolumn{1}{c}{Method} & \multicolumn{1}{c}\qquad\quad\,\,{$a_{\e}$} \qquad\quad& \multicolumn{1}{c}{$a_{\s}$} & \multicolumn{1}{c}{ $b_{\e D}$} & \multicolumn{1}{c}{ $b_{\s D}$} \\
\hline
This work   &\, 6.677424(16)\,&2.6148(3)   & 1.7234(5)  &  0.24757(4)  \\

MC  \cite{Toldin:2021kun}&& 2.60(5)     & &  0.244(8) \\

MC \cite{Przetakiewicz:2025gzi}   &\,  6.679(6) & 2.6143(5)      &  1.69(1) &  0.242(2) \\

FS  \cite{Zhou:2024dbt}&\, 6.4(9)& 2.58(16)    & 1.74(22)  &  0.254(17)\\

CB \cite{Gliozzi:2015qsa} &\,  6.607(7)  & 2.599(1)  & 1.742(6) &   0.25064(6) \\
\hline
\end{tabular}

\bigskip

\begin{tabular}{clllll}
\hline
\hline
\multicolumn{1}{c}{Method} & \multicolumn{1}{c}{$\D_{\s^{\prime}}$} & \multicolumn{1}{c}{ $\D_{\e^{\prime\prime}}$} & \multicolumn{1}{c}{ $\D_{\s^{\prime\prime}}$} \\
\hline
This work   & 5.28901(3)&  6.873(7)  &  8.42915(36)   \\

CB \cite{Simmons-Duffin:2016wlq} & 5.2906(11)& 6.8956(43)  &     \\

CB  \cite{Gliozzi:2015qsa} & 5.49(1) &  7.27(5)& 10.6(3)       \\
\hline
\end{tabular}
\caption{Some 3d Ising data from this work, Monte Carlo simulations \cite{Toldin:2021kun, PTD-10, Hasenbusch-10c,Przetakiewicz:2025gzi},  
fuzzy sphere (FS) regularization \cite{Zhou:2024dbt}, 
and the previous conformal bootstrap studies \cite{Gliozzi:2015qsa,Simmons-Duffin:2016wlq}. 
}\label{res1}
\end{table}

\begin{figure}[t]
    \centering
    \begin{subfigure}[t]{0.40\textwidth}
        \centering
        \includegraphics[width=\textwidth]{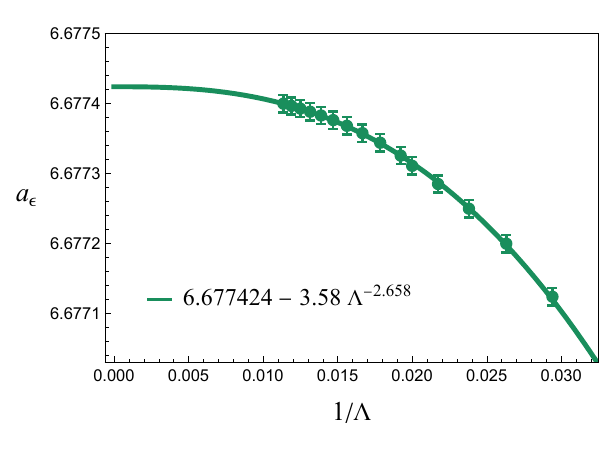}
        \label{aepsilon-N1}
    \end{subfigure}%
    \\
    \begin{subfigure}[t]{0.40\textwidth}
        \centering
        \includegraphics[width=\textwidth]{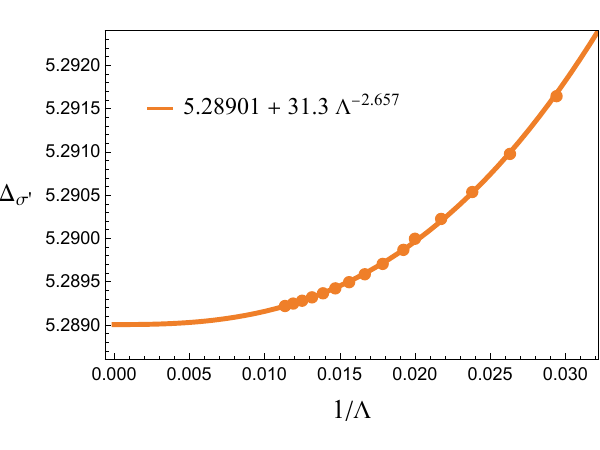}
        \label{deltasigmaprime-N1}
    \end{subfigure}
    \caption{Power-law fits of some 3d Ising results from the $\<\s\,\e\>$ crossing equation. 
    The input-induced errors are indicated by error bars, which are smaller than the plot symbols for $\Delta_{\s'}$. 
}
\label{Conv-se}
\end{figure}

\paragraph{More details about the Ising case.}In Table \ref{res1}, we list some of our highly accurate boundary bootstrap results for the Ising BCFT 
and some literature results for comparison. In Fig. \ref{Conv-se}, we further present the power-law fits of our most accurate results, 
$a_\e$ and $\D_{\s'}$, from the $\<\s\,\e\>$ crossing equation. 
Our prediction for $a_\e$ appears to be 2 orders of magnitude more accurate than the latest Monte Carlo result in \cite{Przetakiewicz:2025gzi}. 
Our estimate for $\D_{\s'}$ is also well consistent with the previous bulk bootstrap result in \cite{Simmons-Duffin:2016wlq} and the more rigorous result, 
$5.262(\textbf{89})$, in \cite{Reehorst:2021hmp}.

\bigskip
\bigskip
 
\onecolumngrid

  \parbox[c][3em][t]{\textwidth}{\centering \large\bf Supplemental Material}
\smallskip

\section{Bulk input parameters for the O($N$) boundary bootstrap}
The explicit input of bulk scaling dimensions for $N=2,3,4,5$ are listed in table \ref{input}. 
The leading O$(N)$ singlet and traceless symmetric tensor are denoted by $S$ and $T$. We use primes to indicate subleading operators in the same representation.

\section{Direct results of the truncated boundary bootstrap}
As the bulk identity is absent in the bulk OPE $\s\times \e$, 
the normalization of the mixed correlator $\<\s(x)\,\e(y)\>$ in the Ising BCFT is not fixed. 
If we normalize the coefficient of the $\s$ conformal block in the bulk channel to unity, 
then the coefficients for the other blocks are rescaled by a factor of $1/(\l_{\s\s\e}\, a_{\s})$. 
In particular, the coefficient of the boundary identity is given by
\be
\frac{b_{\s I}\,b_{\e I}}{\l_{\s\s\e}\, a_{\s}}=\frac{a_{\s}\,a_{\e}}{\l_{\s\s\e}\, a_{\s}}=\frac{a_{\e}}{\l_{\s\s\e}}\,, 
\ee
which can be determined to high accuracy. 
In table \ref{simc}, we list some direct Ising bootstrap results from the two correlators $\<\s(x)\,\e(y)\>$ and $\<\s(x)\,\s(y)\>$. 
Using the above ratio $a_\e/\l_{\s\s\e}$ from $\<\s(x)\,\e(y)\>$ and the product $\l_{\s\s\e}\,a_\e$ from $\<\s(x)\,\s(y)\>$, we can extract the bulk OPE coefficient
\be
\l_{\s\s\e}^{\text{this work}}=\sqrt{\left(\frac{a_{\e}}{\l_{\s\s\e}}\right)^{-1}(\l_{\s\s\e}\, a_{\e})}
=1.05184(13)
\ee 
from our boundary bootstrap results. 
In table \ref{bulkcoefficients-2345}, we also summarize some direct results from the $N>1$ boundary bootstrap. 
The estimates associated with $(\s''', \e''',S'')$ may be less reliable due to mixing effects. 

In addition, we can deduce the Zamolodchikov norm of the displacement operator
\be
C_ D=\left(\frac{\D_{\phi}\,{a_{\phi}}{}}{4\pi\,b_{\phi D}}\right)^2
={0.2368(11)}, 0.2883(16), 0.346(7), 0.3914(32),
\ee
for $N=2, 3, 4, 5$.

\begin{table}\centering\begin{tabular}{ccccc}
\toprule
$N$ & $\D_{\f}$  & $\D_{S}$ & $\D_{S^{\prime}}$  & $\D_{T}$\\
\hline
2 &   0.51908(1)\cite{Hasenbusch:2025yrl}  &  1.51128(5)\cite{Hasenbusch:2025yrl} &  3.789(4)\cite{Hasenbusch:2019jkj}  &  1.23629(\textbf{11})\cite{Chester:2019ifh}\\
\hline
3 &   0.518936(\textbf{67})\cite{Chester:2020iyt}  &   1.5948(2)\cite{Hasenbusch:2020pwj} &  3.759(2)\cite{Hasenbusch:2020pwj}  &   1.20954(\textbf{32})\cite{Chester:2020iyt}\\
\hline
4 &0.51812(4) \cite{Hasenbusch:2021rse}    & 1.66340(35)\cite{Hasenbusch:2021rse}   &  3.755(5)\cite{Hasenbusch:2021rse} &  $1.1864^{+0.0024}_{-0.0034} $\cite{Kos:2013tga}  \\
\hline
5 &0.516985(45)\cite{Hasenbusch:2021rse}   &  1.7182(10)\cite{Hasenbusch:2021rse}  & 3.754(7)\cite{Hasenbusch:2021rse} &   $1.1568(10)$\cite{Kos:2013tga} \\
\hline
\end{tabular}
\caption{Bulk scaling dimensions for $N=2,3,4,5$ from the bulk conformal bootstrap \cite{Chester:2019ifh,Chester:2020iyt,Kos:2013tga} and Monte Carlo simulations \cite{Hasenbusch:2019jkj, Hasenbusch:2020pwj,Hasenbusch:2021rse,Hasenbusch:2025yrl}. 
They are input parameters in our O($N>1$) boundary bootstrap. 
 }
\label{input}
\end{table}

\begin{table}[t]
\begin{tabular}{ccccccc}
\toprule
 $\frac{a_{\e}}{\l_{\s\s\e}}$& $\frac{\l_{\s\s'\e}\,a_{\s'}}{\l_{\s\s\e}\,a_{\s}}$  &$\frac{\l_{\s\s''\e}\,a_{\s''}}{\l_{\s\s\e}\,a_{\s}}$ & \quad$\frac{b_{\e D}\,b_{\s  D}}{\l_{\s\s\e}\,a_{\s}}$ & $\frac{b_{\e \hat{N}}\,b_{\s \hat{N}}}{\l_{\s\s\e}\,a_{\s}}$ & $\frac{b_{\e \hat{N}'}\,b_{\s \hat{N}'}}{\l_{\s\s\e}\,a_{\s}}$ &$\D_{\s'''}$  \\[3pt]
\hline
{6.348244(15)}& $2.260389^{+0.000027}_{-0.000035}$& ${0.3497919^{+0.0000043}_{-0.0000039}}$ &0.1551292(10)& $7.38847^{+0.00069}_{-0.00061}\times10^{-4}$ & ${2.200^{+0.020}_{-0.014}}\times10^{-5}$ &
$11.7708^{+0.0014}_{-0.0019}$  \\
\hline
\end{tabular}\\
\bigskip
\begin{tabular}{cccccccc}
\toprule
 $\l_{\s\s\e}\,a_{\e}$  & $\l_{\s\s\e^{\prime}}\,a_{\e^{\prime}}$ & $\l_{\s\s\e''}\,a_{\e''}$  &$a_{\s}^2$& $b_{\s D}^2$
 &$b_{\s \hat{N}}^2$ & $b_{\s \hat{N}^{\prime}}^2$&$\Delta_{\epsilon'''}$\\
\hline

 7.0235(18)  & 2.25215(59) & 0.19654(24)& 6.8371(12) &  $0.06129(2)$
 &${8.96_{-0.17}^{+0.20}}\times 10^{-5}$ & \quad$1.70_{-0.24}^{+0.33}\times10^{-6}$\quad & ${10.111^{+0.028}_{-0.034}}$  \\
\hline
\end{tabular}\bigskip
\caption{Some direct results of the Ising boundary bootstrap.  
The first table is associated with the spin-energy correlator $\langle \s\,\e\rangle$, while the second table is related to the spin-spin correlator $\langle \s\,\s\rangle$. 
}\label{simc}
\end{table}

\begin{table}\centering\begin{tabular}{ccccccccc}
\toprule
$N$ & $\Delta_{S''}$ & $\l_{\f\f S}\,a_S$  & $\l_{\f\f S'}a_{S'}$ & $\l_{\f\f S''}a_{S''}$  & $\l_{\f\f T}\,a_{T}$ & $\l_{\f\f T'}a_{T'}$ & $b_{\f\hat S}^2$ & $b_{\f\hat V}^2$\\
\hline
2 &${6.63(4)}$&   $3.8278^{+0.0079}_{-0.0070}$  & $1.3122^{+0.0021}_{-0.0023}$ &  $0.1371^{+0.0013}_{-0.0015}$*  &  $4.7283^{+0.0063}_{-0.0056}$& $1.5909^{+0.0025}_{-0.0023}$ &$2.19^{+0.38}_{-0.24}\times10^{-4}$&$6.0^{+1.0}_{-0.8}\times10^{-5}$\\
\hline
3 & ${6.420^{+0.024}_{-0.037}}$&   $2.8148^{+0.0060}_{-0.0051}$  &  ${0.9935^{+0.0019}_{-0.0023}}$ & $0.12067^{+0.00072}_{-0.00070}$ &  ${7.3515^{+0.0085}_{-0.0075}}$&$2.6194^{+0.0051}_{-0.0049}$* & $3.05^{+0.62}_{-0.33}\times10^{-4}$& $4.8^{+1.1}_{-0.9}\times10^{-5}$\\
\hline
4 & ${6.34^{+0.06}_{-0.12}}$ & $2.354^{+0.017}_{-0.010}$    &  ${0.8328_{-0.0046}^{+0.0026}}$   & $0.1109_{-0.0017}^{+0.0022}$ & $9.527^{+0.026}_{-0.023}$& $3.553^{+0.026}_{-0.017}$* & ${4.4^{+2.0}_{-1.5}\times10^{-4}}$* & $5^{+8}_{-3}\times10^{-5}$*\\
\hline
5 & $6.088^{+0.063}_{-0.067}$ &$2.1235^{+0.0094}_{-0.0091}$  & $0.7269^{+0.0041}_{-0.0044}$  & ${0.1087(21)}$* &  $11.497^{+0.035}_{-0.034}$ &$4.481^{+0.021}_{-0.028}$*& ${1.8(3)\times10^{-4}}$* & ${1.4(3)\times10^{-4}}$*\\
\hline
\end{tabular}
\caption{ Some direct results of the O$(N)$ boundary bootstrap for $N=2,3,4,5$. 
Most of them are extracted from power-law fits.   
We use * to label the results from linear fits. }
\label{bulkcoefficients-2345}
\end{table}

\begin{figure}[t]
    \centering
    \begin{subfigure}[t]{0.48\textwidth}
        \centering
        \includegraphics[width=\textwidth]{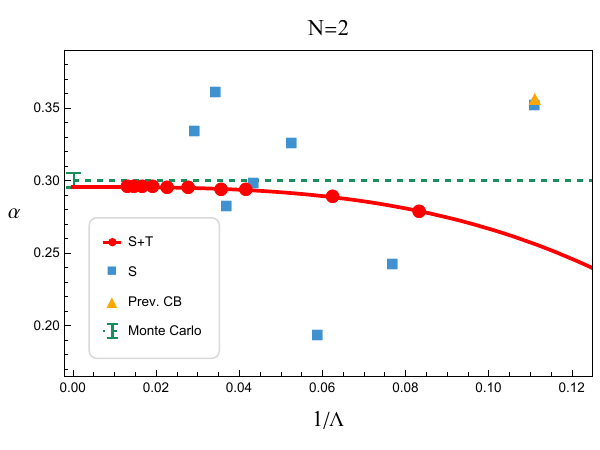}
    \end{subfigure}%
    ~
    \begin{subfigure}[t]{0.48\textwidth}
        \centering
        \includegraphics[width=\textwidth]{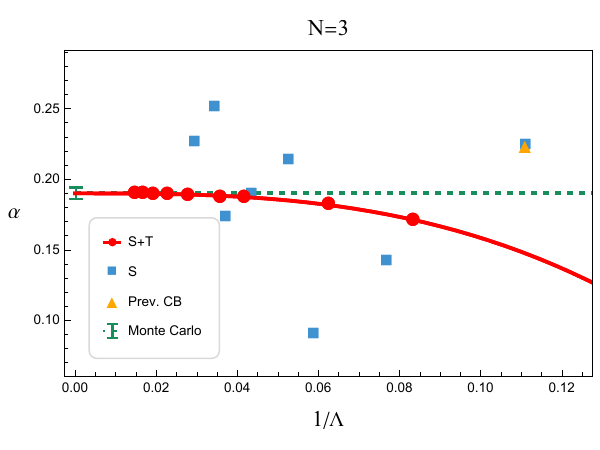}
    \end{subfigure}
     ~
    \begin{subfigure}[t]{0.48\textwidth}
        \centering
        \includegraphics[width=\textwidth]{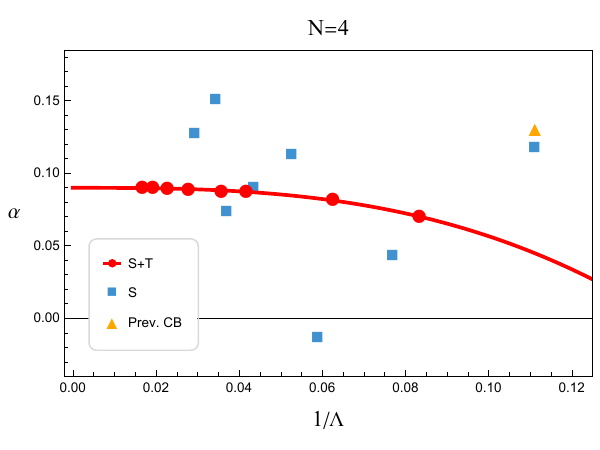}
    \end{subfigure}
     ~
    \begin{subfigure}[t]{0.48\textwidth}
        \centering
        \includegraphics[width=\textwidth]{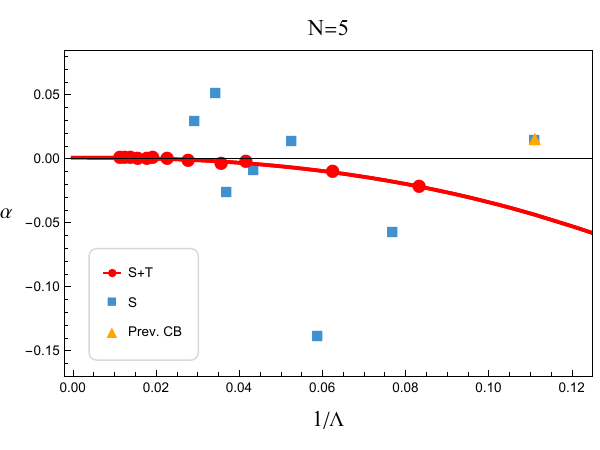}
    \end{subfigure}
    \caption{ The renormalization group parameter $\a$ at different truncation orders $\L$ for $N=2,3,4,5$. 
    The red dots represent the truncated bootstrap solutions with both $S$ and $T$ contributions in the bulk channel. 
    The blue squares are the solutions obtained by projecting the bulk-channel onto the O($N$) singlet $S$. 
    Our $S$ crossing solutions at $1/\L=1/9$ slightly deviate from the previous bootstrap results  \cite{Padayasi:2021sik} (yellow triangles) due to some input differences. 
    The Monte Carlo results \cite{Toldin:2021kun} for $N=2,3$ are represented by green dashed lines with error bars. }
    \label{alpha-solutions-comparison}
\end{figure}

\section{Truncated bootstrap estimates for $\a$ $\text{and $N_c$}$}
For $N>1$, if we project out the traceless-symmetric $T$ contribution and use only one bootstrap equation, 
the truncated solutions do not converge with the truncation order. 
It seems important to balance the numbers of bulk and boundary operators in the truncated bootstrap, 
and thus we take into account the O($N$) traceless-symmetric tensors. 
In Fig. \ref{alpha-solutions-comparison}, we compare the two types of truncated bootstrap solutions for $\a$. 
For $\L=9,13,17,19,23,27,29,34$, the solutions from the crossing equation with only O($N$) singlets $S$ do not appear to converge as $\L$ grows. 
Their deviations exhibit a similar pattern for different $N$, 
where the significant growth of $\a$ at $\L=19, 29$ is due to adding only bulk operators. 
This type of discontinuous behavior also appears in the crossing solutions with both $S$ and $T$ contributions, but is much less significant, 
and thus does not ruin the convergence of our truncated bootstrap solutions. 

\begin{figure}[t]
    \centering
    \includegraphics[width=0.5\textwidth]{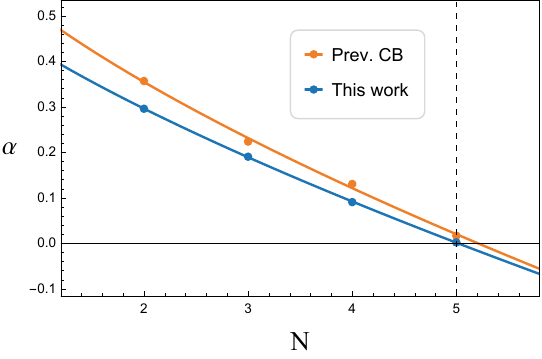}
\caption{
Truncated bootstrap estimates for $\a|_{N=2,3,4,5}$ and their {power-law} fits. 
The blue curve is associated with the best estimates from this work,  
while the orange curve is based on the previous bootstrap estimates in  \cite{Padayasi:2021sik}. 
Our solution of $\a|_{N=N_c}=0$ is $N_c^\text{this work}\approx 5.016$, which is very close to the dashed line ($N=5$). 
For comparison, the critical value from the previous results in \cite{Padayasi:2021sik} is $N_c^\text{previous}\approx 5.207$.}
\label{alpha-N}
\end{figure}

Based on the estimates for $\a|_{N=2,3,4,5}$, 
we further make some power-law fits in $N$, 
using the ansatz $\a=c_0+c_1 N^{c_2}$ with $(c_0, c_1, c_2)$ being free parameters. 
Then we determine the critical value
\be
N_c^\text{this work}= 5.016(28)\,
\ee
by the critical condition $\a|_{N=N_c}=0$. 
To estimate the error, we also perform separate extrapolations for the upper and lower bounds of $\a|_{N=2,3,4,5}$. 
For comparison, a power-law fit of the previous results in \cite{Padayasi:2021sik} gives $N_c^\text{previous}\approx 5.207$. 
In Fig. \ref{alpha-N}, we compare the fitting curves associated with the $\a$ estimates from this work and those from \cite{Padayasi:2021sik}. 
We observe a better fit for our $\a$'s compared to those from \cite{Padayasi:2021sik}, with all data points lying on the fitted curve in our case. 
Above, we also show that our $N=2,3$ estimates for $\a$ are more compatible with the Monte Carlo results.
Therefore, our new estimate for $N_c$ should be more accurate.

According to our improved determination, it seems possible that the critical value $N_c$ is exactly $5$. 
If this is true, there must be an interesting reason behind it,  
and one may establish the exact value of $N_c$ analytically.

\section{Derivation of the truncated bootstrap solutions}
\label{pro}

In this part, we provide some details about the derivation of the truncated bootstrap solutions. 

Let us first explain the notation for the truncation type. We use a pair of integers $(n_\text{bulk},n_\text{bdy})$ to label the truncations. For $N=1$, $(n_\text{bulk},n_\text{bdy})$ indicates the numbers of the bulk and boundary conformal blocks in a bootstrap equation. For $N=2,3,4,5$, the numbers of the bulk O($N$) singlets $S$ and traceless-symmetric tensors  $T$ are both $n_{\text{bulk}}$, 
but there are $n_{\text{bdy}}$ and $n_{\text{bdy}}-1$ boundary operators in O($N-1$) singlet $\hat{S}$ and vector $\hat{V}$ representations, respectively. 
The bulk identity is not counted, but the boundary identity is counted, as the expansion coefficient of the latter is not fixed to unity. 

We minimize the $\eta$ function using \texttt{FindMinimum} in \texttt{Mathematica}, 
which performs a local minimization. 
Since most of the truncated bootstrap solutions are unphysical and their number grows rapidly with the truncation order, 
the local minimization approach is crucial for reaching high  truncation orders $\L$. 
To arrive at a zero of $\eta$, the \texttt{FindMinimum} should start from a well-chosen point,  
i.e., we need to guess a good starting point before knowing the exact location of the zero. 
Our approach is to infer from the solution at a lower truncation order, 
which works nicely. 

To initialize this process, we need to obtain at least one solution at a low truncation order. 
In this case, we can use the homotopy continuation method to deduce all the solutions of 
the truncated bootstrap equations. 
We first reformulate them as a set of polynomial equations 
using the rational approximations of conformal blocks \cite{Lauria:2017wav}. 
As an algebraic geometry problem, we compute the approximate solutions using the efficient package \texttt{HomotopyContinuation.jl}. 
Many solutions are associated with complex numbers. 
Only a few are real and satisfy the physical spectral properties, 
which is a main reason for switching to the local minimization at higher truncation orders. 
We treat a small imaginary part as zero according to the numerical precision. 
We impose some lower bounds for the unknown operator dimensions, 
i.e., they should be greater than the input or protected dimensions, 
$(\D_\s, \D_{\e'}, \D_{S'}, \D_T, \hat \D_t, \hat \D_D)$, according to their representations. 
Unexpectedly, we find at most one physical solution in each truncated crossing system. 
Sometimes we add a small gap to eliminate some unphysical solutions with nearly degenerate low-lying spectrum.  
For instance, we impose a gap $\D_{S''}>\D_{S'}+1/2$ at $\L=12$. 
We conjecture that 
each truncated bootstrap system has at most one physical solution for the normal boundary universality class. 
We carry out this procedure for the truncation order $\L\leq 12$. 

Based on the low truncation solutions, 
we add extra operators to both channels of the crossing equations. 
For $N=1$, we typically add one bulk operator and one boundary operator to a crossing equation at the same time. 
For $N>1$, we usually add a bulk O($N$) singlet, a bulk O($N$) traceless-symmetric tensor, a boundary O($N-1$) singlet, and a boundary O($N-1$) vector, simultaneously. 
However, we may encounter a situation in which 
the maximum scaling dimension in the boundary spectrum is excessively large, 
such as $\hat \D_{n_\text{bdy}}>2\hat \D_{n_\text{bdy}-1}$. 
If this happens, we only add one operator to the bulk channel for $N=1$ 
and two bulk operators for $N>1$.

Then we explain how to construct the starting points. 
As the truncation order increases, we notice that 
the scaling dimensions of low-lying operators change more gently than the high-lying ones, 
which is natural because a physical truncated solution should be interpreted as an effective description. 
Therefore, we do not modify the low-lying spectrum  
and select a series of discrete values for the new and some high dimensions. 
A solution is considered valid if the minimized $\eta$ function is smaller than a threshold set by our numerical precision $\texttt{prec}$, 
e.g., $\eta_{\text{min}}\lesssim 10^{-\texttt{prec}}$. 
In \texttt{FindMinimum}, we usually set a large $\texttt{workingprecision}$, such as $500$.
If no solution is found, we decrease the spacing of the dimensions or take into account 
more high-lying operators. 
To reduce the size of the high-dimension set, we also use another strategy. 
Instead of scanning more starting values of high dimensions, we use the best unsuccessful results as our starting values and scan the same set of high dimensions. 
The starting values of the coefficients of conformal blocks are less important, 
as they are determined by the linear least squares method for a fixed set of scaling dimensions. 
In practice, we find it useful to set the starting values of these coefficients to one. 
As in the cases of low truncation orders, we obtain at most one physical solution in each truncated bootstrap system. 

In the end, we promote the approximate solutions to numerically exact solutions of the truncated systems 
by switching to the exact expressions of the conformal blocks. 

\begin{table}\centering
\begin{tabular}{cc}
\hline
\hline
$N$& Truncations\\
\hline
$\qquad1\, (\s\e)$   & $\quad(18,14),(19,15),(20,16),(21,17),(22,18),(23,19),(24,20),(25,21)$ \\
\hline
$\qquad1\, (\s\s)$  & $(14,10),(15,11),(16,12),(17,13),(18,14),(19,15),(20,16)$ \\
\hline
$2$& $(7,4),(8,5),(9,6),(10,7),(11,8),(12,9)$\\
\hline
$3$& $(7,4),(8,5),(9,6),(10,7),(11,8)$\\
\hline
$4$& $(7,4),(8,5),(9,6),(10,7)$\\
\hline
$5$& $(10,6),(11,7),(12,8),(13,9),(14,10)$\\
\hline
\end{tabular}\caption{The selected truncations for the $\L\rightarrow \infty$ extrapolations. }\label{trun}
\end{table}

\begin{figure}[t]
    \centering
    \begin{subfigure}[t]{0.48\textwidth}
        \centering
        \includegraphics[width=\textwidth]{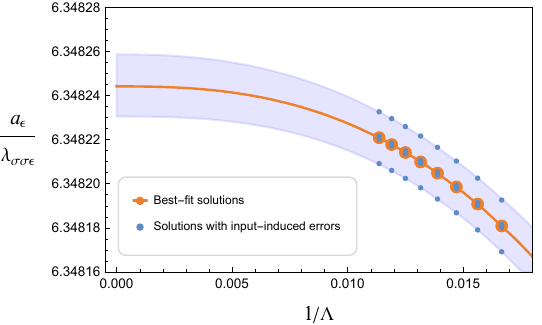}
    \end{subfigure}
    \quad
    \begin{subfigure}[t]{0.48\textwidth}
        \centering
        \includegraphics[width=\textwidth]{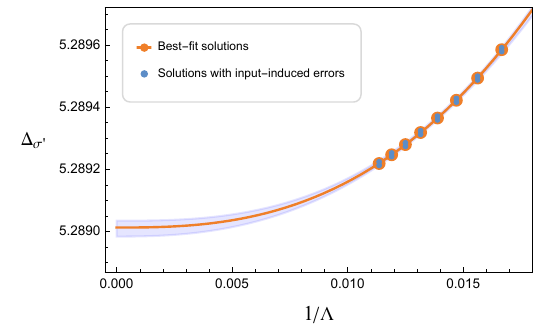}
    \end{subfigure}
    \caption{  Power-law fits of $a_\e/\l_{\s\s\e}$ and $\D_{\s'}$ in $1/\L$ for $N=1$. 
    These results are derived from the $\langle\s\,\e\rangle$ crossing equation. 
    The boundary of a blue band corresponds to the fitting that takes the maximum or minimum value at $\L=\infty$. }
\label{Ext_1}
\end{figure}

\begin{figure}[t]
    \centering
    \begin{subfigure}[t]{0.48\textwidth}
        \centering
        \includegraphics[width=\textwidth]{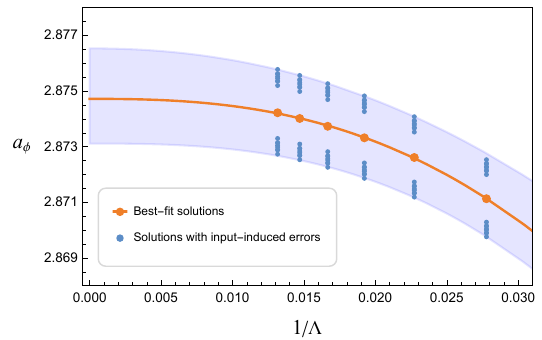}
    \end{subfigure}%
     \quad
    \begin{subfigure}[t]{0.48\textwidth}
        \centering
        \includegraphics[width=\textwidth]{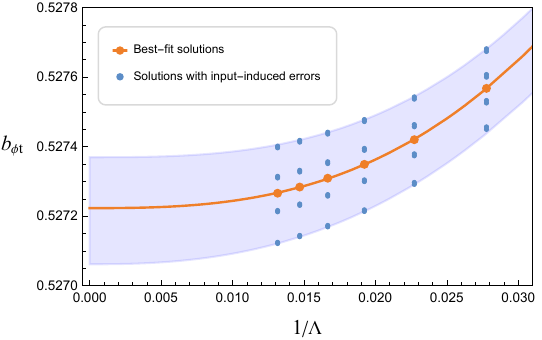}
    \end{subfigure}
    \caption{  Power-law fits of $a_\phi$ and $b_{\phi t}$  in $1/\L$ for $N=2$. 
    The boundary of a blue band is determined by the most distant fittings at $1/\L=0$.}
\label{Ext_2}
\end{figure}

\section{Error analysis}
Below we explain our procedure for estimating errors. 
We assume that the truncated bootstrap results converge to the exact values if the input parameters are exact. 
Therefore, we have two sources of error: finite truncations and input uncertainties. 

To reduce the truncation errors, we use {power-law} fits in $1/\L$ for the truncated solutions 
at the high truncation orders in table \ref{trun}, 
and we extract the $\L\rightarrow \infty $ extrapolations. 
In some cases, the results of the simpler linear fits are more reasonable, such as $\D_{T'}$ and $b^2_{\s N'}$. 
To estimate the extrapolation uncertainties, 
we omit one set of approximate solutions randomly and 
perform extrapolations with the remaining solutions. 
For $\langle\s\e\rangle$, we omit two sets.  
The uncertainty of an extrapolation is determined by the maximum and minimum values at $1/\L=0$.

We now discuss how to estimate the input-induced errors. 
We use the superscripts "+" and "-" to indicate the largest and smallest values 
from the error bars of the bulk scaling dimensions. 
For  $\langle\s\,\s\rangle$, we only consider the input uncertainty from $\D_{\e^{\prime}}$ 
as the uncertainties from $\D_\s, \D_\e$ are much smaller. 
The errors associated with the input uncertainties are deduced from the solutions 
with $\D^+_{\e^{\prime}}$ and $\D^-_{\e^{\prime}}$ in the bulk input. 
For $\langle\s\,\e\rangle$, we consider the errors associated with the uncertainties of $(\D_{\s},\D_{\e})$. 
We solve the truncated bootstrap equations using 4 sets of input choices, 
i.e., $(\D^+_{\s},\D^+_{\e}),(\D^+_{\s},\D^-_{\e}),(\D^-_{\s},\D^+_{\e}),(\D^-_{\s},\D^-_{\e})$. 
For $N=2,3,4,5$, the input uncertainties are associated with $(\D_{\f},\D_{S},\D_{S^{\prime}},\D_{T})$, 
so the number of input choices is $2^4=16$.

Both sources of error are taken into account in our final estimates. 
We extract the truncated solutions associated with different input choices separately. 
Each set of input choice leads to an uncertainty range associated with the $\L\rightarrow \infty$ extrapolation.  
Their maximum and minimum values determine the total errors. 

In Fig. \ref{Ext_1} and Fig. \ref{Ext_2}, 
we present some zoom-in examples for the power-law fits with input-induced and extrapolation errors. 
The uncertainty ranges are indicated by the blue bands.

\end{document}